\documentclass[11pt]{article}

\usepackage[final]{acl}

\usepackage{times}
\usepackage{latexsym}
\usepackage{amsmath}
\usepackage{amssymb}
\usepackage{booktabs}

\usepackage[T1]{fontenc}

\usepackage[utf8]{inputenc}

\usepackage{microtype}

\usepackage{inconsolata}

\usepackage{graphicx}

\title{GoldenRetriever: Non-Interactive Homomorphic Encrypted Retrieval for Privacy-Preserving RAG}

\author{
Yang Gao$^{1}$,
Gang Quan$^{2}$,
Scott Piersall$^{1}$,
Qian Lou$^{1}$,
Dongdong Wang$^{3}$,
Liqiang Wang$^{1}$ \\[0.5ex]
$^{1}$Department of Computer Science, University of Central Florida, Orlando, FL, USA \\
$^{2}$Electrical and Computer Engineering Department,
Florida International University, Miami, FL, USA \\
$^{3}$College of Design, Construction and Planning, University of Florida, Gainesville, FL, USA \\[0.5ex]
\texttt{\{yang.gao, sc382961, qian.lou, liqiang.wang\}@ucf.edu} \\
\texttt{dongdongwang@dcp.ufl.edu} \\
\texttt{gaquan@fiu.edu}
}

\begin{document}
\maketitle
\begin{abstract}
Retrieval-Augmented Generation (RAG) enhances large language models by incorporating external knowledge, but existing pipelines typically operate on plaintext data, raising significant privacy concerns. Prior work on privacy-preserving retrieval leverages cryptographic techniques such as homomorphic encryption (HE) and private information retrieval (PIR), but often relies on interactive protocols or ranking-based selection mechanisms that incur high latency and potential information leakage. 

In this paper, we propose a practical non-interactive encrypted retrieval framework for RAG based on threshold selection. Instead of performing expensive top-$k$ ranking under encryption, our approach selects documents whose similarity scores exceed a predefined threshold, reducing computational complexity from quadratic to linear in the corpus size. We implement this design using CKKS-based homomorphic computation, enabling fully encrypted similarity evaluation and document selection without revealing query content, intermediate scores, or selected indices. 

To bridge the gap between approximate encrypted computation and discrete token reconstruction, we introduce a precision-stable mask polarization method that ensures accurate recovery of selected documents. Experiments on standard retrieval benchmarks demonstrate that our approach achieves competitive retrieval effectiveness while significantly reducing latency compared to ranking-based encrypted methods. 

These results highlight threshold-based selection as a practical foundation for scalable and secure RAG systems.
\end{abstract}

\section{Introduction}

Retrieval-Augmented Generation (RAG) has emerged as a dominant paradigm for knowledge-intensive natural language processing tasks, enabling large language models (LLMs) to incorporate external information during inference. By retrieving relevant documents from a corpus and conditioning generation on them, RAG systems improve factual accuracy and domain adaptability. However, standard RAG pipelines assume that both user queries and document corpora are processed in plaintext on the retrieval server, which raises serious privacy concerns in applications involving sensitive data, such as healthcare, finance, and enterprise knowledge systems.

To address these concerns, recent work has explored privacy-preserving retrieval using cryptographic techniques such as homomorphic encryption (HE), private information retrieval (PIR), and secure multi-party computation. While these approaches enhance data confidentiality, many rely on interactive protocols or multi-round communication between the client and server. Such interaction not only increases latency and system complexity, but also introduces additional avenues for information leakage, including access patterns and intermediate similarity scores, which may expose sensitive relationships between queries and documents. In contrast, practical RAG systems typically favor a simple, stateless retrieval interface, motivating the need for non-interactive encrypted retrieval mechanisms.

A central challenge in this setting is enabling efficient and secure document selection without interaction. Existing secure retrieval methods often depend on iterative querying or adaptive selection procedures, which further exacerbate communication overhead and leakage risks. A natural approach to non-interactive selection is top-$k$ retrieval; however, implementing top-$k$ ranking under homomorphic encryption is computationally prohibitive due to its reliance on large numbers of pairwise comparisons and costly polynomial approximations. Our experiments confirm this limitation: homomorphic top-$k$ selection incurs extremely high latency (e.g., exceeding $10^4$ seconds per query even at modest scales), rendering it impractical for real-world deployment. This exposes a fundamental mismatch between conventional retrieval paradigms and the constraints imposed by encrypted computation.

In this paper, we propose a practical non-interactive encrypted retrieval framework based on \textit{threshold selection} as an alternative to top-$k$ ranking. Instead of explicitly sorting or ranking documents, our method selects documents whose similarity scores exceed a predefined threshold, reducing computational complexity from quadratic to linear in the corpus size. We implement this selection using polynomial indicator functions under the CKKS homomorphic encryption scheme, enabling fully encrypted similarity evaluation and document filtering without interaction. To support downstream generation, we further design a precision-safe token extraction pipeline that ensures accurate reconstruction of selected documents despite the approximate nature of the CKKS homomorphic encryption scheme

We evaluate our framework on multiple retrieval benchmarks and demonstrate that threshold-based encrypted retrieval achieves comparable effectiveness to plaintext baselines while significantly reducing latency compared to top-$k$-based approaches. Our results highlight the necessity of rethinking retrieval design under homomorphic encryption and establish threshold-based selection as a practical foundation for secure RAG systems.

\paragraph{Contributions.}
We summarize our contributions as follows:
\begin{itemize}
    \item We present a novel threshold-based non-interactive encrypted retrieval framework for privacy-preserving RAG, eliminating the need for homomorphic top-k ranking while preserving retrieval quality.

    \item We introduce a precision-stable mask polarization method that enables reliable discrete token reconstruction from approximate CKKS computations.

    \item We show both theoretically and empirically that threshold selection reduces encrypted retrieval complexity from quadratic-ranking style computation to linear document-wise evaluation.

    \item Extensive experiments demonstrate that our approach preserves retrieval effectiveness while substantially reducing latency compared with homomorphic ranking-based retrieval.
\end{itemize}

\section{Related Work}

\paragraph{Privacy-preserving RAG.}
Recent work has begun to explore privacy-preserving retrieval-augmented generation (RAG) systems. 
RemoteRAG~\cite{remoterag2025} studies cloud-based RAG with privacy-aware query processing, combining embedding perturbation and secure distance computation to reduce information leakage. 
SecureRAG~\cite{securerag2025} further considers end-to-end security risks in RAG, including prompt injection and retrieval leakage, and integrates secure retrieval with controlled document access. 
Other systems such as ppRAG~\cite{pprag2025} and p2RAG~\cite{p2rag2026} explore privacy-preserving RAG in outsourced or untrusted environments using combinations of cryptographic primitives and system-level defenses. 
However, these approaches typically rely on multi-stage or interactive retrieval procedures, which introduce additional communication overhead and potential leakage of access patterns.

\paragraph{Secure and encrypted retrieval.}
A large body of work studies privacy-preserving retrieval through private information retrieval (PIR), secure multi-party computation (MPC), and homomorphic encryption (HE). 
These techniques enable hiding query content or retrieval behavior, but many require multiple rounds of interaction or iterative probing~\cite{chor1998private,gentry2009fully}. 
Recent work on encrypted dense retrieval extends these ideas to embedding-based search, allowing similarity computation over encrypted vectors~\cite{mazzone2025efficient}. 
While effective for secure ranking, these methods typically focus on retrieval itself and do not address the downstream requirement of reconstructing discrete token sequences for generation.

\paragraph{Homomorphic ranking and approximate computation.}
Homomorphic comparison, sorting, and top-$k$ selection under CKKS have been studied using polynomial approximation techniques~\cite{ckks2017,mazzone2025efficient}. 
These methods enable ranking over encrypted data, but the approximate nature of CKKS introduces numerical errors that lead to near-binary outputs rather than exact decisions. 
Such approximation is often acceptable for ranking metrics, but it becomes problematic when subsequent operations require discrete correctness, such as token extraction in RAG pipelines.

\paragraph{Our contribution.}
In contrast to prior work, we study a complementary design point: a \emph{non-interactive encrypted retrieval framework} for RAG, where a retrieval server processes a single encrypted query and returns encrypted results in one shot without interaction. 
Furthermore, we explicitly address the gap between approximate homomorphic selection and discrete token reconstruction, ensuring reliable document extraction for downstream generation.

\section{Problem Setup and Threat Model}

We consider a three-party architecture consisting of a client, a retrieval server, and an LLM server. The document corpus is stored in plaintext on the retrieval server. The LLM server generates a public/secret key pair under the CKKS homomorphic encryption scheme, along with the required evaluation keys (e.g., relinearization and rotation keys). The public key is distributed to both the client and the retrieval server, while the evaluation keys are provided to the retrieval server to enable homomorphic computation. The secret key remains exclusively on the LLM server. The client encodes a query into an embedding, encrypts it using the public key, and submits the encrypted query to the retrieval server.

The retrieval server is assumed to be \emph{honest-but-curious}: it follows the prescribed protocol but may attempt to infer sensitive information from observable data. Since the server does not possess the secret key, it cannot decrypt the query, intermediate similarity scores, or final outputs. All similarity computation and document selection are performed in the encrypted domain using homomorphic operations.

To enable non-interactive retrieval, the server applies a threshold-based selection mechanism over encrypted similarity scores to identify relevant documents, without iterative querying or multi-round communication. The server then performs masked token extraction and returns an encrypted representation of the candidate set to the LLM server. In this representation, selected documents preserve their original token values, while non-selected documents are masked to zero within the ciphertext. Consequently, the retrieval server gains no information about which documents are selected.

The LLM server decrypts the received ciphertexts, reconstructs the selected documents, and performs downstream generation. Because the full corpus structure is preserved and no indices are revealed during retrieval, our framework prevents access-pattern leakage.

Our design ensures that the retrieval server does not observe the query content, similarity scores, or selected document indices. We emphasize that our threat model protects query privacy and selection privacy, while the document corpus itself remains in plaintext on the retrieval server.

Our objective is to enable accurate RAG-style retrieval under strong privacy constraints, ensuring (1) query confidentiality, (2) protection of intermediate similarity scores, and (3) concealment of selected document indices, all within a non-interactive retrieval pipeline.

Our framework is designed for deployment scenarios where the document corpus is owned by or intentionally exposed to the retrieval service, while user queries and retrieval outcomes remain confidential. Protecting the document corpus itself is outside the scope of this work.

\section{Method}

We propose a non-interactive encrypted retrieval framework for retrieval-augmented generation (RAG). Our approach performs similarity computation, threshold-based selection, and document extraction entirely in the encrypted domain, without revealing query content or selection results to the retrieval server. In the following subsections, we describe each component in detail.

\begin{figure*}[t]
    \centering
    \includegraphics[width=\textwidth]{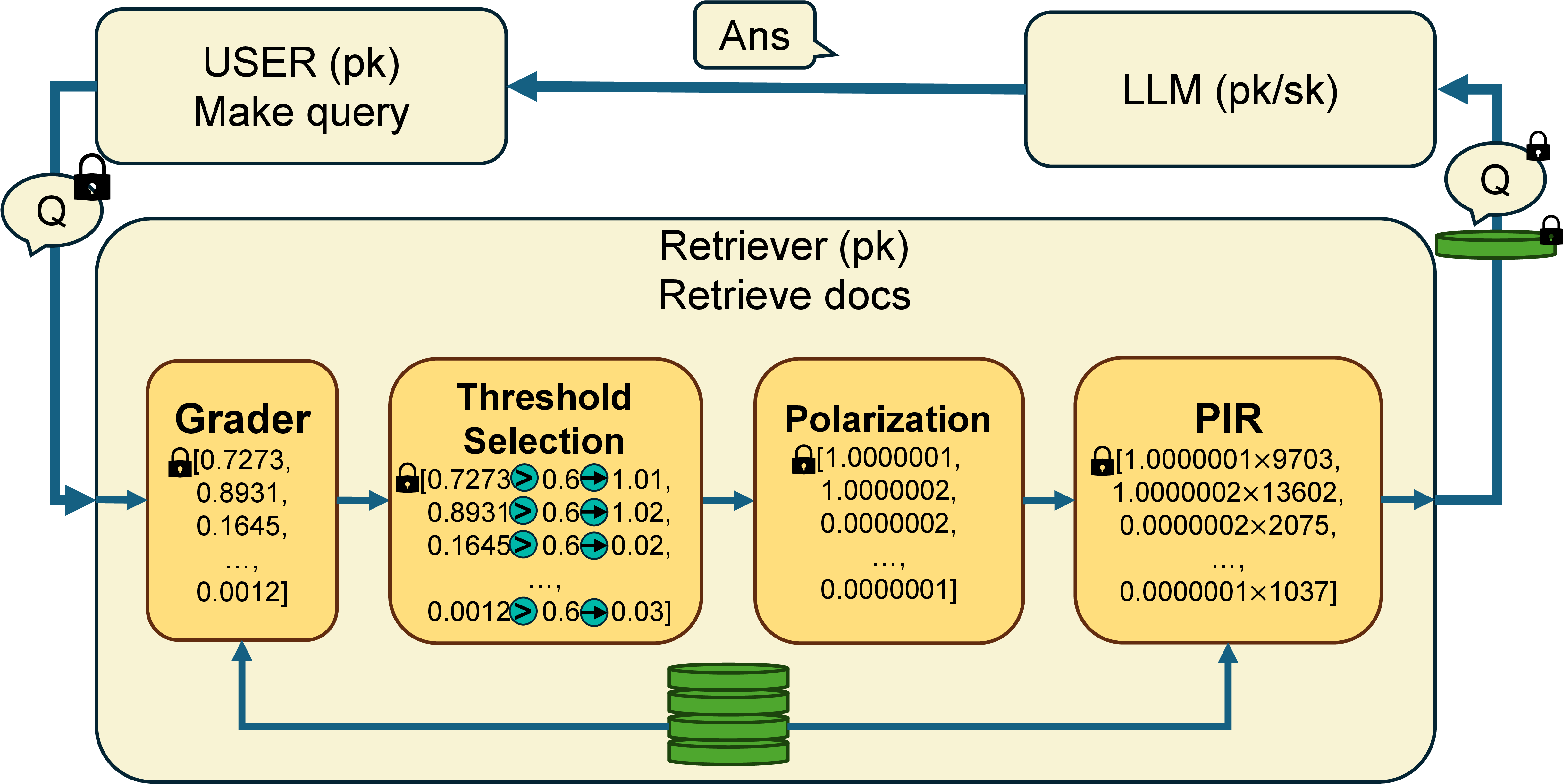}
    \caption{Overview of the proposed non-interactive encrypted retrieval framework for privacy-preserving RAG.}
    \label{fig:framework}
\end{figure*}

\subsection{Architecture Overview}

Our framework enables non-interactive encrypted retrieval for RAG through a three-party architecture consisting of a client, a retrieval server, and an LLM server. The design ensures that encrypted retrieval integrates seamlessly with downstream generation while preserving query and selection confidentiality.

\subsubsection{Key Generation and Query Encryption.}
The LLM server generates a public–secret key pair under the CKKS homomorphic encryption scheme, along with evaluation keys (e.g., relinearization and rotation keys). The public key is distributed to the client, while the evaluation keys are provided to the retrieval server. The secret key remains exclusively on the LLM server. Upon receiving a user query, the client encodes it into a normalized embedding vector $\mathbf{q}\in\mathbb{R}^D$, encrypts it using the public key, and submits the encrypted query to the retrieval server.

\subsubsection{Encrypted Retrieval.}
The retrieval server stores the document corpus in plaintext form. Each document is represented in two formats:
\begin{itemize}
    \item \textbf{Embedding matrix} $\mathbf{D}\in\mathbb{R}^{B\times D}$, where each row $\mathbf{d}_i$ is a normalized embedding of document $i$.
    \item \textbf{Token matrix} $\mathbf{X}\in\mathbb{Z}^{B\times T}$, where each row contains the fixed-length token ID sequence of document $i$.
\end{itemize}
These two matrices share the same row index, ensuring that similarity computation and token extraction remain consistent.

Given the encrypted query embedding, the retrieval server computes encrypted cosine similarity scores for all documents, applies threshold-based selection using a homomorphic indicator function, and produces an encrypted binary mask indicating selected documents. All selection operations are executed without decryption and without interaction with the client.

\subsubsection{Encrypted Document Extraction and Generation.}
Using the encrypted selection mask, the retrieval server performs masked token extraction over the document matrix and returns encrypted token sequences corresponding to selected documents. These ciphertexts are sent to the LLM server, which decrypts them using the secret key and proceeds with standard RAG-style generation.

This separation ensures that (1) the retrieval server never observes plaintext queries or selected document indices, (2) the client never accesses the secret key, and (3) the LLM server only receives selected documents after encrypted filtering.

A central technical challenge lies in ensuring that approximate homomorphic computation produces selection masks sufficiently precise for discrete token extraction, which we address in Section 4.4.

\subsection{Encrypted Cosine Similarity}
We adopt embedding-based dense retrieval using cosine similarity. All document embeddings are precomputed, normalized, and stored in plaintext on the retrieval server. Let
\begin{equation}
    \mathbf{D}\in\mathbb{R}^{B\times D}
\end{equation}
denote the embedding matrix, where each row $\mathbf{d}_i$ is unit-normalized.

Given a normalized query embedding $\mathbf{q}\in\mathbb{R}^{D}$, the client encrypts $\mathbf{q}$ using CKKS and submits the ciphertext to the retrieval server. Cosine similarity reduces to a dot product:
\begin{equation}
    s_i = \mathbf{q}^{\mathsf T}\mathbf{d}_i
\end{equation}

Because document embeddings are stored in plaintext, similarity computation is performed as ciphertext–plaintext multiplications followed by homomorphic additions. This design significantly reduces computational depth and noise growth compared to ciphertext–ciphertext operations.

To enable efficient batched evaluation over $B$ documents, we leverage CKKS SIMD packing. The encrypted query vector is replicated across slots corresponding to document embeddings, allowing parallel evaluation of multiple similarity scores within a single ciphertext. The resulting encrypted score vector
\begin{equation}
    \mathbf{s} \in \mathbb{R}^B
\end{equation}
contains approximate similarity values for all documents and serves as input to the threshold-based selection mechanism.

\subsection{Threshold-Based Selection}

Given the encrypted similarity score vector
\[
\mathbf{s} \in \mathbb{R}^B,
\]
where each element $s_i$ represents the similarity between the query and document $i$, our goal is to identify relevant documents in a non-interactive manner.

\paragraph{Selection via Thresholding.}

We adopt a threshold-based selection strategy, where documents with similarity scores exceeding a predefined threshold $\tau$ are selected. The selection mask is defined as
\begin{equation}
m_i = \mathbf{1}(s_i > \tau),
\end{equation}
where $\mathbf{1}(\cdot)$ denotes the indicator function.

This formulation avoids explicit ranking and enables document selection using only pointwise operations over the score vector.

\paragraph{Homomorphic Implementation.}

Under homomorphic encryption, the indicator function cannot be evaluated directly. We adopt the indicator function construction introduced in~\cite{mazzone2025efficient}, which enables evaluating range conditions over encrypted values using polynomial-based techniques.

Specifically, given a threshold $\tau$, we compute
\begin{equation}
m_i \approx \mathrm{Ind}(s_i - \tau),
\end{equation}
where $\mathrm{Ind}(\cdot)$ is an indicator function that outputs values close to $1$ when the input lies within a target interval and close to $0$ otherwise. In practice, this is implemented in CKKS as a Chebyshev-approximated interval indicator function, following~\cite{mazzone2025efficient}, where the target rank range is encoded as a homomorphic range test.

\paragraph{Non-Interactive Property.}

The threshold-based formulation allows all documents to be evaluated independently without iterative refinement or multi-round interaction. This enables a fully non-interactive retrieval process, where the server computes selection results in a single pass over the encrypted score vector.

\paragraph{Discussion.}

Unlike fixed-size selection mechanisms, threshold-based selection produces a variable number of retrieved documents. In practice, the threshold $\tau$ can be tuned to balance retrieval quality and output size. We study the impact of this trade-off in Section~6.

\subsection{Precision-Stable Mask Polarization}

Threshold-based selection under homomorphic encryption produces an approximate binary mask
\begin{equation}
m = b + \epsilon, \quad b \in \{0,1\}, \quad |\epsilon| \le \delta,
\end{equation}
where $b$ denotes the ideal binary selection result and $\epsilon$ captures approximation errors introduced by CKKS encoding, rescaling, and polynomial evaluation of the indicator function.

\paragraph{Precision Requirement.}

After masking a token ID $x \in [0,V)$, we obtain
\begin{equation}
\tilde{x} = m x.
\end{equation}

To guarantee correct integer reconstruction after decryption and rounding, we require
\begin{equation}
|\tilde{x} - x| < 0.5.
\end{equation}

Assuming a conservative vocabulary upper bound $V = 60{,}000$, this imposes
\begin{equation}
|m - 1| < \frac{0.5}{V}
= 8.33 \times 10^{-7}.
\end{equation}

A symmetric bound applies when $m$ should equal $0$. Therefore, the polarization function must reduce mask deviation below this tolerance.

\paragraph{Polynomial Construction.}

We construct a polynomial $f(x)$ satisfying the following boundary constraints:

\begin{align}
f(0) &= 0, & f(1) &= 1, \\
f'(0) &= 0, & f'(1) &= 0, \\
f''(0) &= 0, & f''(1) &= 0, \\
f'''(0) &= 0, & f'''(1) &= 0.
\end{align}

These eight independent constraints require a polynomial of degree at least seven. Solving the resulting linear system yields the minimal-degree solution:
\begin{equation}
f(x) = 35x^4 - 84x^5 + 70x^6 - 20x^7.
\end{equation}

This polynomial exhibits third-order flatness at both $0$ and $1$, ensuring strong contraction of small perturbations near binary values.

\paragraph{Error Contraction Analysis.}

For $m = b + \epsilon$ with $b \in \{0,1\}$ and small $\epsilon$, Taylor expansion gives
\begin{equation}
f(b+\epsilon)
= b + \frac{f^{(4)}(b)}{4!}\epsilon^4 + O(\epsilon^5).
\end{equation}

Therefore, the output error scales as
\begin{equation}
|f(m) - b| = O(\epsilon^4).
\end{equation}

Given empirical mask deviation $\delta \approx 0.02$, we obtain
\begin{equation}
\delta^4 \approx 1.6 \times 10^{-7},
\end{equation}
which satisfies the required tolerance up to a moderate constant factor. This confirms that the chosen degree-7 polynomial achieves sufficient precision for discrete token recovery.

\paragraph{HE Cost Consideration and Generalization.}

The degree of the polarization polynomial is determined by the required output precision, which in turn depends on the vocabulary size $V$. Specifically, to guarantee correct token recovery after rounding, the mask error must satisfy
\begin{equation}
|m - b| < \frac{0.5}{V}.
\end{equation}
Given an upper bound on the input deviation $\delta$, the polynomial degree can be chosen such that the contracted error $O(\delta^r)$ meets this requirement, where $r$ is determined by the order of flatness at the boundaries.

In our setting with $V = 60{,}000$ and empirical $\delta \approx 0.02$, a degree-7 polynomial (with third-order flatness) is sufficient. For different vocabulary sizes or precision requirements, the same construction procedure can be applied to select an appropriate polynomial degree. Larger vocabularies require stricter output precision and may necessitate higher-degree polynomials, while smaller vocabularies allow lower-degree approximations.

The degree-7 polynomial can be evaluated using Horner-style computation with three multiplicative levels under CKKS, maintaining manageable noise growth. Increasing the degree improves error contraction but incurs additional multiplicative depth and rescaling cost, highlighting a trade-off between precision and efficiency.

\subsection{Secure Token Extraction}

After obtaining the polarized mask vector
\[
\mathbf{m} \in \{0,1\}^B,
\]
we extract selected documents in the encrypted domain without revealing their indices.

Recall that the corpus is represented as a token matrix
\[
\mathbf{X} \in \mathbb{Z}^{B \times T},
\]
where each row corresponds to a document and each column corresponds to a token position.

\paragraph{Column-wise Masking.}

For each token position $j \in \{1, \dots, T\}$, we construct the column vector
\[
\mathbf{x}_j = \mathbf{X}_{:,j} \in \mathbb{Z}^B.
\]

We then compute the masked column
\begin{equation}
\tilde{\mathbf{x}}_j = \mathbf{m} \odot \mathbf{x}_j,
\end{equation}
where $\odot$ denotes element-wise multiplication. Since $\mathbf{m}$ is encrypted and $\mathbf{x}_j$ is stored in plaintext, this operation corresponds to ciphertext–plaintext multiplication under CKKS.

The resulting $\tilde{\mathbf{x}}_j$ remains encrypted and contains nonzero entries only for selected documents.

\paragraph{Full-Structure Return.}

To prevent index leakage and access-pattern exposure, we return all $B$ rows of the masked matrix to the LLM server, including rows corresponding to non-selected documents (which evaluate to zero after masking). The retrieval server does not observe which rows are effectively selected.

Formally, the retrieval server outputs encrypted columns
\[
\{\tilde{\mathbf{x}}_1, \dots, \tilde{\mathbf{x}}_T\},
\]
which together represent the masked document matrix.

\paragraph{Decryption and Reconstruction.}

Upon receiving the encrypted masked matrix, the LLM server decrypts each column and reconstructs token sequences. Due to the precision guarantees established in Section~4.4, masked token values satisfy
\[
|\tilde{x} - x| < 0.5,
\]
ensuring correct integer recovery via rounding.

\paragraph{Security Considerations.}

Because masking and extraction occur entirely in the encrypted domain, the retrieval server does not learn the query content, similarity scores, or selected document indices. Returning the full $B \times T$ structure further prevents access-pattern leakage during retrieval.

\section{Complexity and Security Analysis}
We analyze the computational cost of each component in our encrypted retrieval pipeline to understand its scalability under homomorphic encryption. Our framework is designed to avoid iterative interaction and ranking procedures, enabling a single-pass evaluation over the document corpus. As a result, the overall complexity is linear in the number of documents, with the dominant cost arising from similarity computation and token extraction. Below, we break down the complexity of each stage.

\subsection{Computational Complexity}

Let $B$ denote the number of documents, $D$ the embedding dimension, and $T$ the fixed token length per document.

\paragraph{Similarity Computation.}

Encrypted cosine similarity requires $O(BD)$ ciphertext--plaintext multiplications and homomorphic additions. Since document embeddings are stored in plaintext, this stage avoids ciphertext--ciphertext multiplication, significantly reducing multiplicative depth and noise growth. By leveraging CKKS SIMD packing, multiple similarity scores can be computed in parallel within a single ciphertext.

\paragraph{Threshold-Based Selection.}

Threshold-based selection operates independently on each similarity score, resulting in $O(B)$ homomorphic evaluations. Each selection is implemented via a polynomial-based indicator function, requiring only a small number of multiplicative levels. This design avoids sorting or pairwise comparisons and enables efficient single-pass selection over the encrypted score vector.

\paragraph{Mask Polarization.}

Mask polarization applies a fixed-degree polynomial to each element of the selection mask, resulting in $O(B)$ operations. In our implementation, the degree-7 polynomial requires three multiplicative levels and introduces only modest overhead compared to similarity computation.

\paragraph{Token Extraction.}

Masked token extraction requires $O(BT)$ ciphertext--plaintext multiplications, as each column of the token matrix is multiplied with the encrypted mask vector. This stage dominates the overall computational cost due to the size of the document matrix.

\paragraph{Overall Complexity.}

The total computational complexity of our framework is
\[
O(BD + B + BT) = O(B(D + T)),
\]
which scales linearly with the corpus size. In practice, the dominant cost arises from similarity computation and token extraction. In our current implementation, end-to-end latency for $B=20$ documents is on the order of $10^3$ seconds per query. While slower than plaintext retrieval, this cost reflects a fully non-interactive encrypted pipeline with strong privacy guarantees.

Future optimizations may reduce latency through improved batching strategies, more efficient packing schemes, or hardware acceleration.

\subsection{Security Analysis}

We assume an honest-but-curious retrieval server that follows the protocol but may attempt to infer sensitive information from observable data.

\paragraph{Query Confidentiality.}

Query confidentiality follows directly from the semantic security of the CKKS encryption scheme. Since the retrieval server only receives ciphertexts encrypted under the public key and never possesses the secret key, it cannot recover the plaintext query embedding beyond negligible probability.

\paragraph{Score Confidentiality.}

All similarity scores and intermediate values remain encrypted throughout computation. The retrieval server observes only ciphertexts and cannot recover plaintext similarity values.

\paragraph{Selection Privacy and Access-Pattern Protection.}

Unlike conventional top-$k$ retrieval, our framework never reveals document indices during encrypted computation. Document selection is performed entirely in the encrypted domain, and the retrieval server always returns a masked representation of the complete candidate set rather than only the selected documents. Selected documents retain their token values, while non-selected documents are masked to zero. Because the output preserves the full candidate-set structure and exposes no explicit indices, the retrieval server cannot distinguish which documents satisfy the retrieval criterion, thereby mitigating access-pattern leakage.

\paragraph{Security Assumptions}
Our analysis assumes the semantic security of the CKKS encryption scheme and an honest-but-curious retrieval server. We do not consider side-channel attacks or malicious adversaries deviating from the protocol.

\paragraph{Limitations.}

The document corpus is stored in plaintext on the retrieval server and is therefore not protected. Our framework focuses on protecting query privacy and selection privacy, and does not aim to conceal the corpus itself.

\section{Experiments}

\subsection{Experimental Setup}

\paragraph{Datasets.}
We evaluate the proposed framework on several retrieval benchmarks, including \textit{MS MARCO}, \textit{Natural Questions}, \textit{HotpotQA}, and \textit{FiQA}. For each dataset, we build evaluation subsets consisting of $Q$ queries and their corresponding candidate document sets.

\paragraph{Embedding Model.}
We use the BGE-base dense retrieval model to encode both queries and documents into 384-dimensional normalized embeddings.

\paragraph{Corpus Construction.}
For each query, we construct a candidate set containing two relevant (positive) documents and a collection of non-relevant (negative) documents. Negative documents are sampled from documents associated with other queries to provide a diverse yet controlled retrieval setting.

Unless otherwise specified, the main experiments use $Q = 50$ queries, each with $B = 100$ candidate documents. To study scalability, we further increase the candidate set size to $B = 500$ and $B = 1000$ while keeping the number of queries fixed.

Because homomorphic computation remains expensive, this controlled construction allows us to systematically evaluate retrieval effectiveness and runtime under encrypted settings.

\paragraph{Token Representation.}
Each document is tokenized into a fixed-length sequence of length $T$, where token IDs are represented as integers.

\paragraph{Implementation Details.}
All experiments were conducted on a server equipped with dual Intel Xeon Silver 4114 CPUs (20 physical cores, 40 hardware threads, 2.20 GHz), 251 GB RAM, running Ubuntu 24.04.4 LTS. The implementation was built using OpenFHE 1.1.2. No GPU acceleration was used.

\paragraph{Evaluation Metrics.}
We report the following metrics:
\begin{itemize}
    \item \textbf{Recall}: fraction of relevant documents successfully selected.
    \item \textbf{Token Accuracy}: percentage of correctly recovered tokens after decryption.
    \item \textbf{Document Accuracy}: percentage of documents whose full token sequences are correctly reconstructed.
    \item \textbf{Latency}: end-to-end runtime per query.
    \item \textbf{Avg. Selected Documents}: average number of documents selected per query.
\end{itemize}

Due to the high computational cost of homomorphic evaluation, we adopt a controlled evaluation protocol following common practice in homomorphic-encryption research. Rather than evaluating over the full benchmark corpora, we construct candidate sets with controlled sizes, allowing systematic comparison of retrieval effectiveness, reconstruction accuracy, and computational cost under different encrypted retrieval strategies. We further evaluate scalability by increasing the candidate set size from 100 to 1000 documents. This controlled setting isolates the effect of the retrieval algorithm from the overwhelming computational cost of large-scale homomorphic inference, enabling reproducible and meaningful comparisons across methods.

\subsection{Main Results}

We compare our threshold-based encrypted retrieval framework against a plaintext baseline and a ranking-based encrypted baseline.

\begin{table*}[t]
\centering
\caption{Main results comparing retrieval effectiveness and efficiency.}
\begin{tabular}{lccccc}
\toprule
Method & Recall & Token Acc (\%) & Doc Acc (\%) & Latency (s) & Avg. Docs \\
\midrule
Plaintext & 1.0 & 100 & 100 & 3.7 & 2 \\
Threshold (Ours) & 1.0 & 100 & 100 & 1051.8 & 2 \\
Ranking-based (HE) & 1.0 & 100 & 100 & 16579.9 & 2 \\
\bottomrule
\end{tabular}
\end{table*}

As shown in Table~1, the proposed threshold-based method achieves the same retrieval and reconstruction accuracy as the plaintext and ranking-based encrypted baselines on this evaluation setting. However, compared with ranking-based encrypted retrieval, our method substantially reduces end-to-end latency, decreasing runtime from 16579.9 seconds to 1051.8 seconds. Although plaintext retrieval remains much faster, the proposed method provides a practical non-interactive encrypted alternative while preserving full reconstruction quality in our experiments.

\subsection{Latency Breakdown}

To better understand the computational bottlenecks of the proposed pipeline, we report the runtime contribution of each major stage.

\begin{table}[t]
\centering
\caption{Latency breakdown of the proposed method.}
\begin{tabular}{lcc}
\toprule
Stage & Time (s) & Percentage \\
\midrule
Similarity Computation & 335.47 & 31.9\% \\
Threshold Selection & 139.22 & 13.2\% \\
Mask Polarization & 38.54 & 3.7\% \\
Token Decryption & 262.88 & 24.5\% \\
Token Reconstruction & 264.92 & 25.2\% \\
Others & 10.80 & 1.0\% \\
\midrule
Total & 1051.8 & 100\% \\
\bottomrule
\end{tabular}
\end{table}

The latency breakdown shows that similarity computation is the single largest component, accounting for 31.9\% of the total runtime. Token decryption and token reconstruction together contribute nearly half of the overall cost, indicating that document recovery remains a major bottleneck in the end-to-end pipeline. In contrast, threshold selection and mask polarization introduce relatively modest overhead, which supports the efficiency advantage of threshold-based selection over ranking-based encrypted retrieval.

\subsection{Scaling Analysis}

We next evaluate scalability by varying the number of candidate documents $B$.

\begin{table}[t]
\centering
\caption{Scaling behavior with respect to corpus size.}
\begin{tabular}{lccc}
\toprule
$B$ & Latency (s) & Recall & Token Acc \\
\midrule
100 & 1051.8 & 1.0 & 100 \\
500 & 2112.7 & 1.0 & 100 \\
1000 & 4361.2 & 1.0 & 100 \\
\bottomrule
\end{tabular}
\end{table}

The results show that retrieval quality remains stable as the corpus size increases, with recall and token accuracy unchanged across all tested settings. At the same time, latency grows with the number of candidate documents, which is consistent with the increasing cost of similarity evaluation and token extraction over larger candidate sets. This trend suggests that the proposed framework scales predictably while maintaining correct document recovery.

\subsection{Effect of Threshold}

We further study the impact of the threshold parameter $\tau$ on retrieval behavior.

\begin{table}[t]
\centering
\caption{Effect of threshold on retrieval performance.}
\begin{tabular}{lccc}
\toprule
$\tau$ & Recall & Avg. Docs & Latency \\
\midrule
Low (0.2) & 1.0 & 2 & 1051.8 \\
Medium (0.5) & 1.0 & 2 & 1051.8 \\
High (0.7) & 0.5 & 1 & 1051.8 \\
\bottomrule
\end{tabular}
\end{table}

Table~4 shows that lower and medium thresholds both preserve full recall while selecting, on average, two documents per query. In contrast, a higher threshold reduces the average number of selected documents to one, but also lowers recall to 0.5, indicating that some relevant documents are filtered out. These results illustrate the trade-off controlled by $\tau$: a more aggressive threshold can reduce output size, but may harm retrieval effectiveness if set too high.

\subsection{Discussion}

Overall, the experimental results demonstrate that threshold-based encrypted retrieval provides an effective compromise between privacy and efficiency. Compared with plaintext retrieval, the encrypted pipeline incurs substantial computational overhead, as expected. However, compared with ranking-based encrypted selection, the proposed method significantly reduces runtime while maintaining perfect token and document reconstruction in our evaluation. These results suggest that threshold-based selection is a practical design choice for non-interactive privacy-preserving retrieval in RAG systems.

\bibliography{custom}
\end{document}